\documentclass[%
 aip,
 amsmath,amssymb,
 reprint,%
]{revtex4-1}

\usepackage{graphicx}
\usepackage{dcolumn}
\usepackage{bm}
\usepackage{enumitem} 
\usepackage[utf8]{inputenc}
\usepackage[T1]{fontenc}
\usepackage{mathptmx}
\usepackage{etoolbox}

\newcommand{\HZO}{Hf$_{1-x}$Zr$_x$O\textsubscript{2}}
\newcommand{\PR}{$P_R$}
\newcommand{\ULK}{$U_{LK}$}
\newcommand{\Uint}{$U_{int}$}
\newcommand{\Ugrad}{$U_{grad}$}
\newcommand{\Uelec}{$U_{elec}$}
\newcommand{\Utot}{$U_{tot}$}
\newcommand{\Vapp}{$V_{app}$}
\newcommand{\omp}{\textit{o}(--)-phase}
\newcommand{\opp}{\textit{o}(+)-phase}
\makeatletter
\def\@email#1#2{%
 \endgroup
 \patchcmd{\titleblock@produce}
  {\frontmatter@RRAPformat}
  {\frontmatter@RRAPformat{\produce@RRAP{*#1\href{mailto:#2}{#2}}}\frontmatter@RRAPformat}
  {}{}
}%
\makeatother
\begin{document}

\preprint{AIP/123-QED}

\title[Zr Concentration-Dependent Sub-Lattice Phase-Field Model of \HZO: Analysis of Phase Composition and Polarization Switching]{Zr Concentration-Dependent Sub-Lattice Phase-Field Model of \HZO: Analysis of Phase Composition and Polarization Switching}
\author{Tae Ryong Kim}
 \email{kim3804@purdue.edu}
\author{Sumeet K. Gupta}%
\affiliation{ 
Elmore Family School of Electrical and Computer Engineering, Purdue University, West Lafayette, Indiana 47906, USA
}%

\date{\today}

\begin{abstract}
We develop a sub-lattice phase-field model of \HZO~incorporating zirconium (Zr) concentration ($x$)-dependence.
Our framework expands the time-dependent Ginzburg-Landau (TDGL) equation to the sub-lattice level and incorporates $x$-dependent interaction parameters and gradient coefficients.
Our experimentally calibrated model captures the evolution of charge-voltage ($Q-V$) characteristics for $x$ ranging from 0.5 to 1.0.
The sub-lattice formulation explains the thermodynamic preference and kinetic transition barriers of competing orthorhombic phase (\textit{o}-phase) and tetragonal phase (\textit{t}-phase), while the phase-field framework enables spatially resolved analysis of polarization ($P$) and electric-field (\textit{E}-field) profiles, allowing multi-domain (MD) $P$ and mixed-phase states to emerge naturally.
Our model reproduces the experimentally observed ferroelectric (FE)-to-anti-ferroelectric (AFE) transition as $x$ increases from 0.5 to 1.0.
At low Zr concentration ($x$ = 0.5--0.6), the o-phase dominates, yielding distinct FE behavior. 
At high concentration ($x$ $\geq$ 0.9), the \textit{t-}phase is stabilized, leading to AFE transitions.
A key finding of our work is the unique behavior at intermediate Zr concentrations ($x$ = 0.7--0.8).
Here, the \textit{o-} and \textit{t-}phase energies are comparable, making the system strongly influenced by local variations in the electric field (\textit{E}-field), which arise from stray fields near the domain walls. 
This non-uniform field distribution results in a mixed-phase composition and spatially staggered $P$ reversal, which manifests as a more gradual $Q-V$ evolution (compared to other values of $x$).
By linking energy landscapes to spatial field effects, the model provides insights into the FE-to-AFE crossover in \HZO.
\end{abstract}

\maketitle

\section{\label{sec:level1}Introduction}

Zirconium-doped hafnium oxide (\HZO) has attracted tremendous attention for its ferroelectric (FE) characteristics along with CMOS compatibility \cite{muller_ferroelectricity_2012}.
With an Hf:Zr ratio 1:1 ($x$ = 0.5), \HZO~exhibits robust ferroelectricity at annealing temperatures of 400--500 \textdegree C. 
This facilitates the integration of Zr-doped HfO\textsubscript{2} with CMOS front-end-of-line at a low thermal budget \cite{muller_ferroelectricity_2012}, making it attractive for the design of various memory devices such as ferroelectric capacitors (FERAM) \cite{okuno_1t1c_2022,okuno_highly_2023}, ferroelectric field effect transistors (FEFETs) \cite{dunkel_fefet_2017,dutta_logic_2022} and ferroelectric tunnel junctions (FTJs) \cite{ryu_ferroelectric_2019,luo_highly_2020}.

Interestingly, the polarization-voltage ($P-V$) response of \HZO~evolves from FE to anti-FE (AFE) behavior as $x$ increases \cite{muller_ferroelectricity_2012,hyukpark_surface_2017,park_morphotropic_2018,jung_review_2022}, thereby enhancing its versatility for various applications \cite{chang_anti-ferroelectric_2020,zhang_ferroelectricantiferroelectric_2025,ravikumar_first_2025}.
At $x \sim$ 0.5, \HZO~shows robust FE switching with high remanent polarization (\PR) ($\sim$ 20 $\mu$C/cm$^{2}$).
Conversely, Zr-rich \HZO~($x \sim$ 1.0) exhibits AFE double-loop hysteresis with near-zero \PR \cite{muller_ferroelectricity_2012}.
The AFE behavior manifests as a reversible field-induced transition, where the material transforms into a polar state under bias and reverts to a non-polar state upon field removal.
This tunability arises from the progressive stabilization of the non-polar tetragonal phase (\textit{t}-phase) (P4\textsubscript{2}/mmc) over the polar orthorhombic phase (\textit{o}-phase) (Pca2\textsubscript{1}) at the spontaneous state \cite{reyes-lillo_antiferroelectricity_2014,materlik_origin_2015}.
In the FE regime, switching in \HZO~is a direct polarization ($P$) reversal within the \textit{o-}phase under non-zero electric-field (\textit{E}-field) \cite{qi_polarization_2025}.
In contrast, the AFE behavior is driven by a reversible field-induced phase transition between \textit{o-} and \textit{t-}phases, as shown in first-principles-based \cite{reyes-lillo_antiferroelectricity_2014} and experimental \cite{lomenzo_discovery_2023} studies.
Here, the application of a non-zero \textit{E}-field causes the phase transition from the stable \textit{t}-phase into the \textit{o}-phase.
Upon removal of the field, the system reverts to the stable \textit{t-}phase.

To understand such phase transitions, the works in \cite{saha_microscopic_2019} modeled $x$-dependent charge-voltage ($Q-V$) characteristics using a sub-lattice model for the fluorite \HZO~lattice.
The model explains phase stabilization in terms of the interaction energy (\Uint) and gradient energy (\Ugrad) between the two sub-lattices.
Here, the \textit{o}-phase possesses parallel $P$ for each sub-lattice, whereas the \textit{t}-phase has an anti-parallel sub-lattice $P$, resulting in a net-zero averaged $P$ for a single lattice.
If \Uint~and \Ugrad~impose an energy cost for any deviation from the parallel $P$, \textit{o}-phase becomes stabilized across the \HZO~system.
On the contrary, the \textit{t}-phase is stabilized when anti-parallel $P$ alignment is energetically favorable in the spontaneous state.
The work \cite{saha_microscopic_2019} suggests that adding Zr to \HZO~impacts \Uint~and \Ugrad~by providing cell compression in the internal oxygen (O) atoms in each sub-lattice \cite{clima_first-principles_2018}.
Due to this compression, adding Zr significantly increases the dipole-dipole interaction, making the anti-parallel O-atom displacement favorable; thereby, \textit{t}-phase becomes the stable over \textit{o}-phase at high Zr concentration.

Despite this capability, this sub-lattice model bears a critical limitation in that it assumes the same phase of the entire lattice across the \HZO~layer.
Unlike the assumption in the sub-lattice model, FE-doped-HfO\textsubscript {2} displays mixed phase within the single grain \cite{grimley_atomic_2018}.
Specifically, around $x \sim$ 0.7, \HZO~exhibits the coexistence of both phases\cite{ni_equivalent_2019}.
To predict the mixed-phase behavior, a multi-domain (MD) phase-field framework \cite{chang_multi-domain_2022}, based on Kittel’s model \cite{kittel_theory_1951}, was proposed that can reproduce FE/AFE and dielectric (DE) responses.
Despite the good agreement of the model with a wide range of measured $Q-V$ loops, its phase composition is prescribed by the user rather than emerging from an $x$-dependent energetic preference \cite{materlik_origin_2015,park_morphotropic_2018}.
Consequently, it does not explain how or why specific \textit{o-} and \textit{t-}phase distributions arise as a function of $x$ and applied voltage.
\begin{figure}[t]
\centering
\includegraphics[width=\columnwidth]{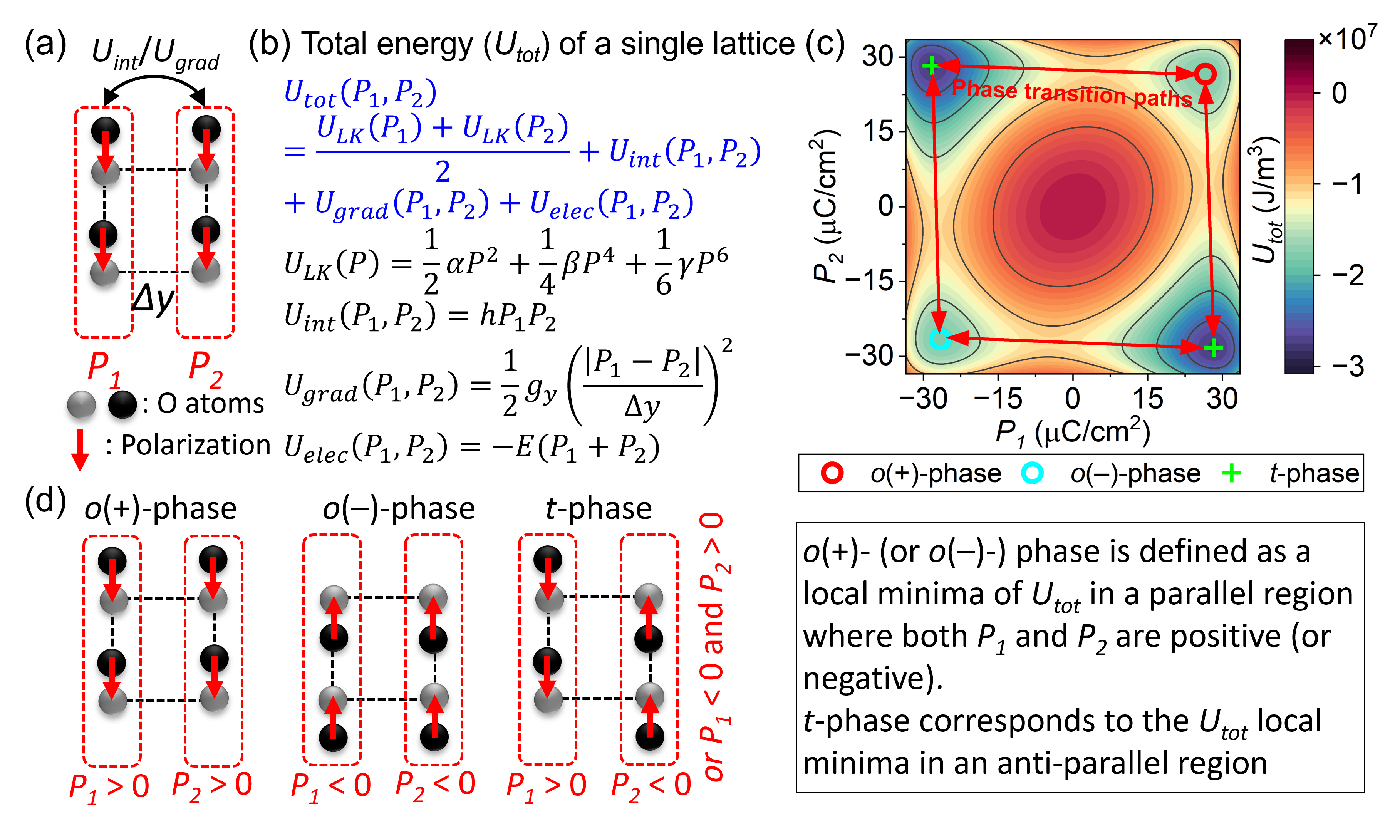}
\caption{\label{fig:epsart} (a) Microscopic schematics of \textit{o-} and \textit{t}-phase at the sub-lattice level. $P_1$ (and $P_2$) stands for the polarization ($P$) of left (right) sub-lattice and $\Delta y$ is the distance between two sub-lattices (b) the equations for the single lattice energy (\Utot) and its components (c) \Utot~versus $P_1$ and $P_2$ map with phase transition paths (four red arrows) (identical to each other due to the symmetry of \Utot) and the definition of the \textit{o}(--)-, \textit{t-}, and \opp~in the map and (d) the microscopic schematics of the each phase.}
\end{figure}

To address these limitations, we propose a self-consistent sub-lattice-based phase-field model. 
We incorporate $x$-dependent sub-lattice interaction energy terms with our phase-field framework, which offers spatial resolution of \textit{o-} and \textit{t}-phases naturally arising from the energetics.
In our model, a single \HZO~layer consists of multiple sub-lattices, which possess spontaneous positive and negative $P$.
Between each sub-lattice, $x$-dependent \Uint~and \Ugrad~are defined to reflect the effect of Zr concentration on the energy between two sub-lattices.
To explain phase stabilization and spatial $P$ switching characteristics as a function of $x$, we analyze the energy landscape from thermodynamic and kinetic perspectives.
The key contributions of this work are summarized below.

\begin{itemize} [leftmargin=*,topsep=0.5ex,itemsep=0.5ex]
\item We propose a self-consistent sub-lattice phase-field model that predictively captures macroscopic $Q-V$ evolution as a function of $x$.
\item Using the sub-lattice energy landscape, we elucidate how the interplay between $x$-dependent single lattice energy and \textit{E}-field governs the $P$ switching characteristics for different $x$.
\item We identify the physical origin of the mixed-phase composition and gradual $P$ switching at intermediate $x$ (= 0.7--0.8), attributing it to comparable energy of the \textit{o-} and \textit{t}-phase and spatially non-uniform \textit{E}-field induced by irregular $P$ domain patterns.
\end{itemize}

\begin{figure}[t]
\centering
\includegraphics[width=\columnwidth]{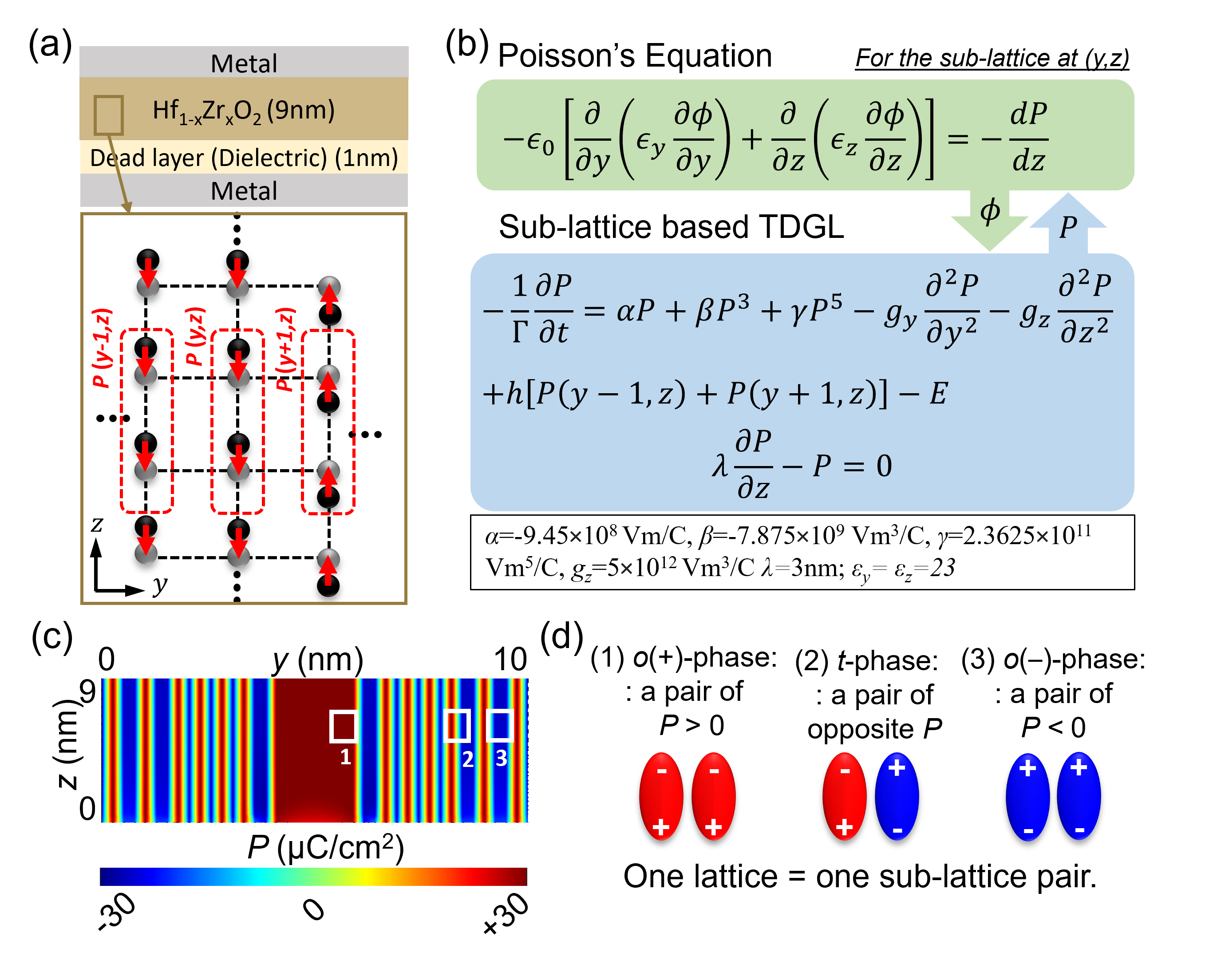}
\caption{\label{fig:epsart} (a) The schematic of \HZO~based MFIM capacitor (top) and microscopic structure of \HZO~at the sub-lattice level (bottom) (b) a set of the equations for sub-lattice based phase-field model framework and $x$-independent parameters (the black lined box) (c) a spatial $P$ map of simulated \HZO~domain and (d) sub-lattice $P$ configurations of \textit{o}(+)-, \textit{t-}, and \omp}
\end{figure}
\begin{figure*}[!t]
\includegraphics[width=\textwidth]{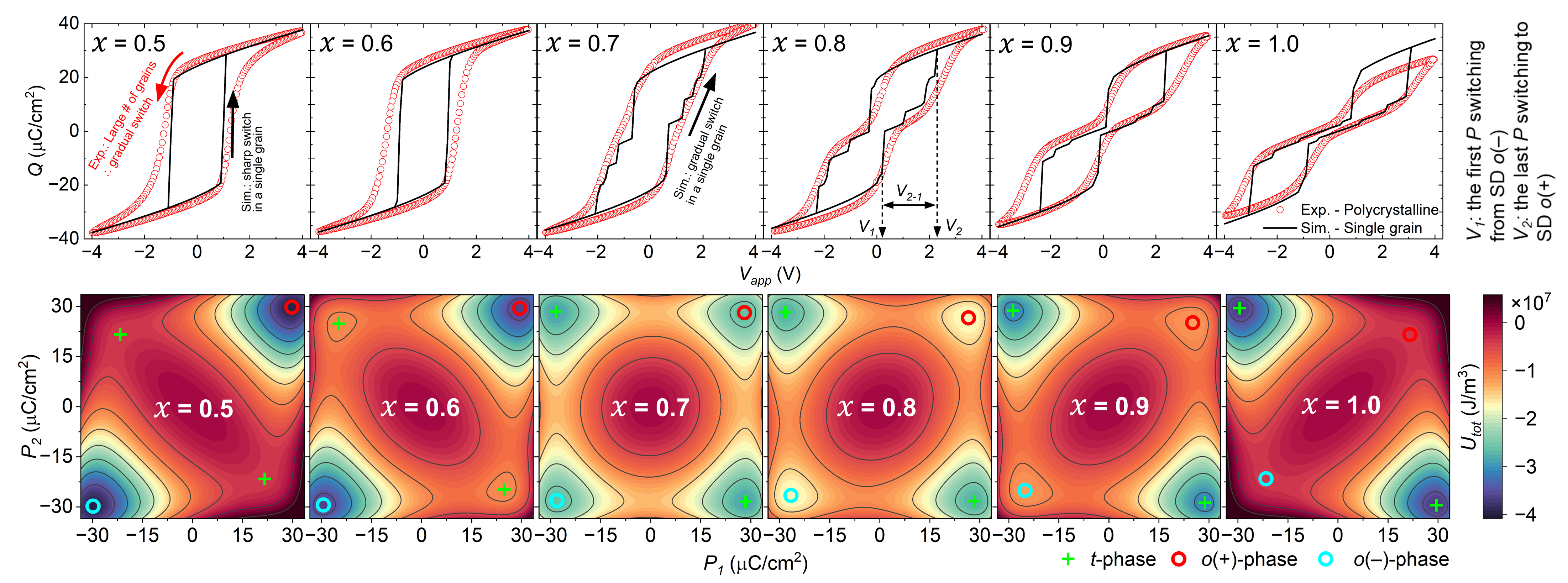}
\caption{(a) Calibration of model (line) for different Zr concentrations $x$ with experimental data (symbols) from \cite{saha_microscopic_2019} (reused with permission) (b) Corresponding computed single-lattice \Utot~ landscapes in the $(P_1, P_2)$ space for $x =$ 0.5--1.0 at $E$ = 0, illustrating the evolution of local minima associated with the \textit{o}(--)-, \textit{t-}, and \textit{o}(+)-phases. The measured curves in (a) are smoother because the device response reflects averaging over many grains. SD \textit{o}(--) (SD \textit{o}(+)) denotes a single-domain state in which the entire grain is in the \omp~(\opp) configuration.}
\label{fig3}
\end{figure*}
\section{Sub-lattice Based Phase-field Model for \HZO Layer}
The sub-lattice model \cite{saha_microscopic_2019} provides a microscopic energy-based interpretation of the observed $x$-dependent $Q-V$ characteristics.
In the model, a single lattice is divided into two sub-lattices, whose $P$ are denoted as $P_1$ (left) and $P_2$ (right).
Each sub-lattice is governed by a local Landau–Khalatnikov free energy \ULK~as described in (1),
\begin{equation}
    \label{eq:placeholder_label}
    U_{LK}(P)=\frac{\alpha}{2}{P^{2}}+\frac{\beta}{4}{P^{4}}+\frac{\gamma}{6}{P^{6}},
\end{equation}
where $\alpha$, $\beta$, and $\gamma$ are the Landau free energy coefficient.
Due to the double-well potential landscape of \ULK~with an energy barrier between, \ULK~gives rise to a non-zero $P$ at the spontaneous state of each sub-lattice \cite{saha_microscopic_2019,saha_negative_2021,chandra_landau_2007}.

Between the two sub-lattices, two different types of $x$-dependent interaction energies exist, namely, the gradient energy (\Ugrad) and the interaction energy (\Uint).
\Ugrad~is characterized by the gradient coefficient ($g_y$) \cite{li_phase-field_2001}.
\begin{equation}
    \label{eq:placeholder_label}
    U_{grad}(P_1,P_2)=\frac{1}{2}g_y\left(\frac{P_1-P_2}{\Delta{y}}\right)^2
\end{equation}
According to the expression for \Ugrad, any difference between the sub-lattice $P$ increases \Ugrad, thereby making it energetically favorable for a lattice to exhibit parallel alignment of $P_1$ and $P_2$.
\Uint~is based on the Kittel-model framework\cite{kittel_theory_1951}, governed by the interaction coefficient ($h$) \cite{chang_multi-domain_2022,hoffmann_antiferroelectric_2022}.
\begin{equation}
    \label{eq:placeholder_label}
    U_{int}(P_1,P_2)=hP_1P_2
\end{equation}
\Uint~influences the $P$ alignment such that positive $h$ ($h >$ 0) energetically favors opposite $P$ directions between sub-lattices, whereas negative $h$ ($h <$ 0) favors parallel $P$ states.
In addition to these internal interaction terms, an external \textit{E}-field (denoted as $E$) generates electrostatic energy ($U_{elec}$), coupled with each sub-lattice $P$.
\begin{equation}
    \label{eq:placeholder_label}
    U_{elec}(P_1,P_2)=-E(P_1+P_2)
\end{equation}

The total free energy of a single lattice \Utot~is derived by summing the averaged \ULK~of left and right sub-lattices, \Ugrad, \Uint, and \Uelec.
\begin{equation}
    \label{eq:placeholder_label}
    U_{tot}(P_1,P_2)=\frac{U_{LK}(P_1)+U_{LK}(P_2)}{2}+U_{grad}+U_{int}+U_{elec}
\end{equation}

Fig. 1(c) shows the 2-D contour map of the \Utot~at the spontaneous ($E$ = 0) as a function of $P_1$ and $P_2$.
Each crystallographic phase is identified with a stable point of the single-lattice free-energy landscape, i.e., local minima of \Utot~in the $(P_1, P_2)$ space \cite{saha_microscopic_2019}.
The polar \textit{o}-phase corresponds to minima with parallel sub-lattice $P$ ($P_1$ and $P_2$ have the same sign), yielding a non-zero net $P$.
We label these minima as \textit{o}(+)-phase for $P_1 > 0$ and $P_2 > 0$, and \textit{o}(--)-phase for $P_1 < 0$ and $P_2 < 0$.
The \textit{t}-phase corresponds to minima with anti-parallel sub-lattice $P$ ($P_1P_2 < 0$), resulting in near-zero net $P$.

Although the sub-lattice model successfully captures the $x$-dependent energetics of a single lattice, a spatially resolved model is still required because practical \HZO~layers comprise a dense network of sub-lattices \cite{grimley_atomic_2018,paul_formation_2024}.
To this end, we develop the phase-field model based on the aforementioned sub-lattice energy terms.
The sub-lattice phase-field framework, which views the \HZO~layer as a grid of sub-lattices (Fig. 2(a)), self-consistently solves the time-dependent Ginzburg-Landau (TDGL) equation and Poisson's equations for each sub-lattice at a given applied voltage to calculate the $P$ and potential ($\phi$) (Fig. 2(b)).
The framework models the metal-FE \HZO-metal (MFM) with a dead layer at the bottom.
We consider the top dead layer to be negligible compared to the bottom one in an \HZO-based MFM capacitor based on \cite{oh_effect_2020,pesic_physical_2016}.
In our framework, the dead layer is a DE layer \cite{pesic_physical_2016,paul_oxygen_2025}, which tends to hold zero $P$ at the spontaneous state.
\begin{figure}[t]
\centerline{\includegraphics[width=\columnwidth]{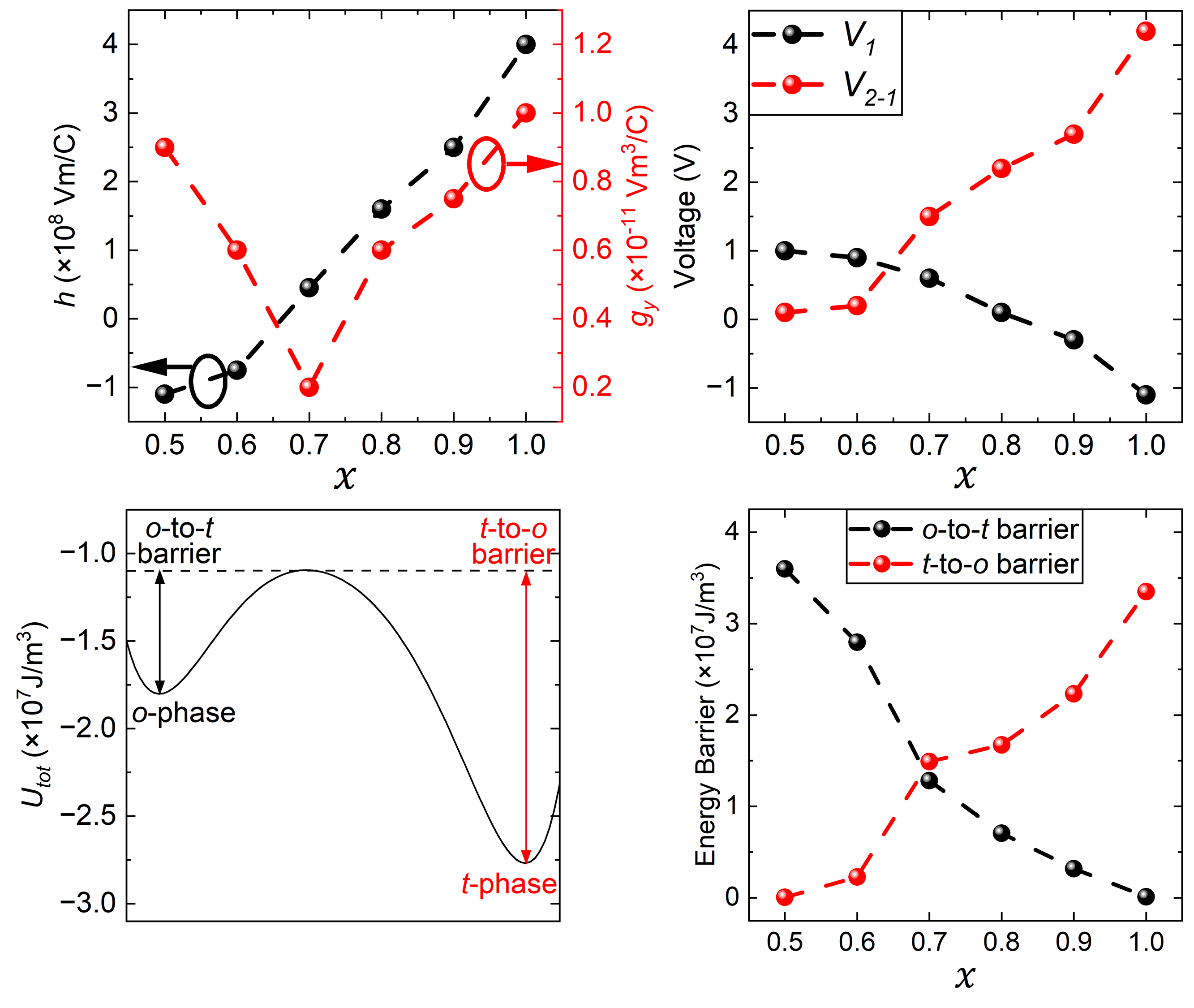}}
\caption{(a) The calibrated $h$ and $g_y$ for different $x$ in the sub-lattice phase-field model (b) The trend of $V_1$ (black) and $V_{2-1}$ (red) versus $x$. (c) 1-D projected phase transition path illustrated in Fig. 1(c). (d) The trend of \textit{o}-to-\textit{t} (black) and \textit{t}-to-\textit{o} (red) energy barriers versus $x$.}
\label{fig4}
\end{figure}
\begin{figure*}[t]
\includegraphics[width=\textwidth]{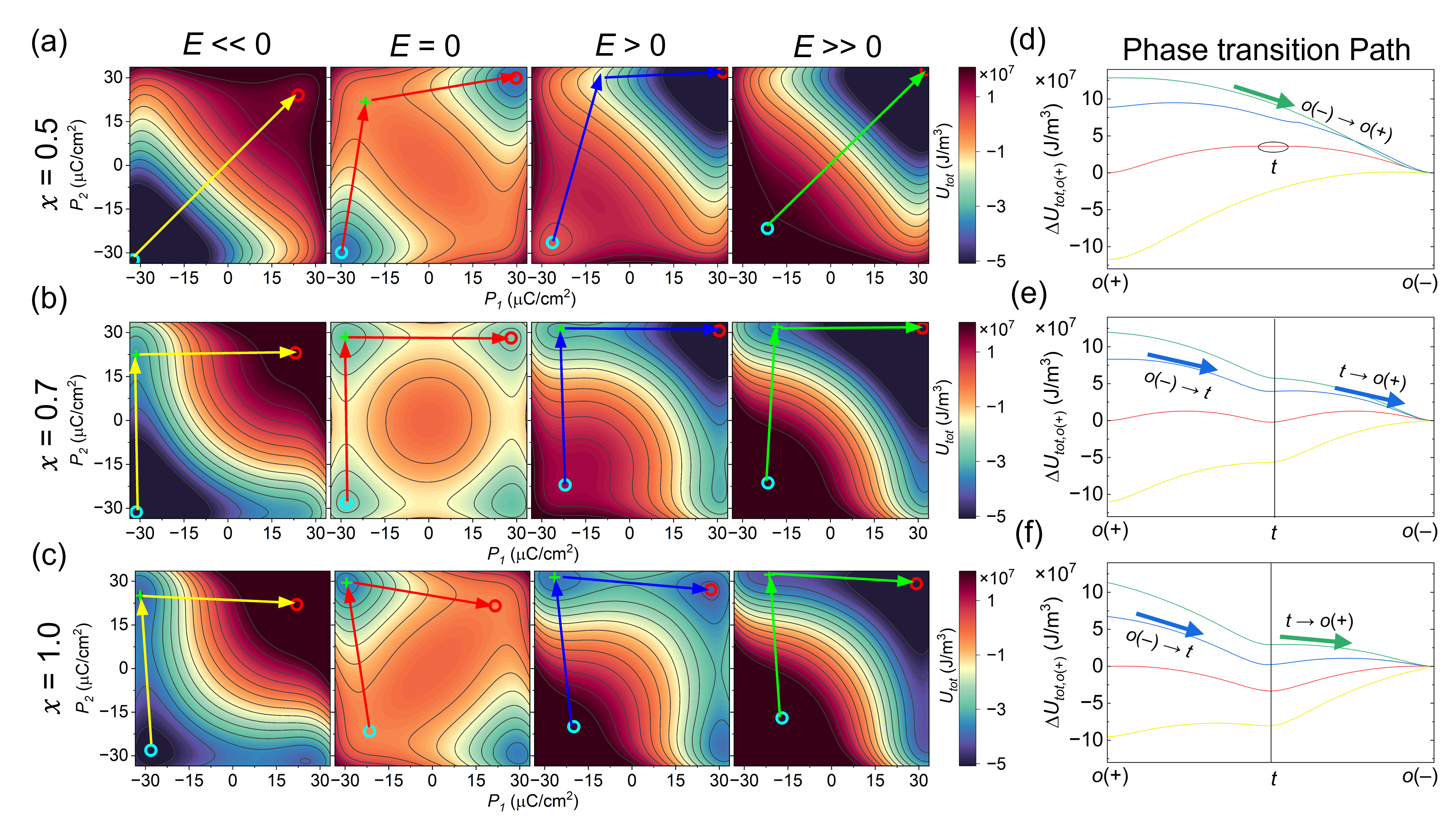}
\caption{Evolution of \Utot~as a function of $P_1$ and $P_2$ during a positive \textit{E}-field sweep (left to right) for $x$ = (a) 0.5, (b) 0.7, and (c) 1.0, and 1-D projections of the phase transition paths at different $E$ (yellow, red, blue, and green arrows in Fig.~5(a)–(c)) for $x$ = (d) 0.5, (e) 0.7, and (f) 1.0. For the 1-D projected paths, $\Delta U_{tot,\textit{o}(+)} = U_{tot} - U_{tot,\textit{o}(+)}$ is defined with respect to the \Utot~of \opp~for better visualization.}
\label{fig5}
\end{figure*}
The TDGL equation for FE \HZO~ layer relates the rate of $P$ change rate to the total energy (\textit{F}) of each sub-lattice to capture the time-dependent behavior of $P$ in the FE layer.
Considering ($y$,$z$) coordinates for the 2D cross section of the FE layer and considering $P$ along the $z$-direction, the TDGL equation relating $P$ change rate and \textit{F} is
\begin{equation}
    \label{eq:placeholder_label}
    -\frac{1}{\Gamma}\frac{dP(y,z)}{d t} = \frac{d F(y,z)}{d P(y,z)},
\end{equation}
where $\Gamma$ is the viscosity coefficient.
\textit{F}($y$,$z$) consists of free ($f_{free}$), gradient ($f_{grad}$), sub-lattice interaction ($f_{int}$) and  electrostatic ($f_{elec}$) energy components, which are described as
\begin{equation}\label{eq:placeholder_label}
    \begin{gathered}
        f_{free}(y,z)=\frac{\alpha}{2}{P(y,z)^{2}}+\frac{\beta}{4}{P(y,z)^{4}}+\frac{\gamma}{6}{P(y,z)^{6}},\\
        f_{grad}(y,z)=\frac{g_{y}}{2}\left(\frac{\partial{P(y,z)}}{\partial{y}}\right)^2
                     +\frac{g_{z}}{2}\left(\frac{\partial{P(y,z)}}{\partial{z}}\right)^2,\\
        f_{int}(y,z) = hP(y,z)[P(y-1,z)+P(y+1,z)],\\
        f_{elec}(y,z)= -E\cdot{P(y,z)}
    \end{gathered}
\end{equation}
where $\alpha$, $\beta$, $\gamma$, $g_y$ and $h$ are the same as the coefficients described before for the sub-lattice model.
Note, we also incorporate the gradient coefficient (\textit{g\textsubscript{z}}) along the FE thickness ($z$-direction), which, recall, is also the direction along which the polarization points. At the FE-DE interface, the surface energy is considered by using,
\begin{equation}\label{eq:placeholder_label}
    \begin{gathered}
        \lambda\frac{\partial{P(y,z)}}{\partial{z}}-P(y,z)=0,\\
    \end{gathered}
\end{equation}
where $\lambda$ (nm) is the screening length \cite{koduru_phase-field_2023}.

Besides the TDGL equation, Poisson's equation captures the electrostatic behavior of the entire MFM stack,
\begin{equation}
    -\varepsilon_0 \left[ \frac{\partial}{\partial y} \left( \varepsilon_y \frac{\partial \phi}{\partial y} \right)
    + \frac{\partial}{\partial z} \left( \varepsilon_z \frac{\partial \phi}{\partial z} \right)\right]
    = -\frac{\partial P(y,z)}{\partial z},
    \label{eq:placeholder_label}
\end{equation}
where $\varepsilon_y$ and $\varepsilon_z$ are the permittivity of \HZO~in $y$ and $z$-direction, respectively (here, we assume the same permittivity for both directions). 

We simulate a single grain FE in this work with its \textit{c}-axis oriented along the $z$ direction.
We consider 9 nm of \HZO~and 1 nm of dead layer (total thickness = 10 nm) in our simulations, which is consistent with \cite{hsain_wake-up_2023}.
The simulation domain dimension corresponds to the average size of a single grain of \HZO~film with 10 nm thickness \cite{hyukpark_surface_2017}.
Zr concentration ($x$)-dependent $h$ and $g_y$ parameters are calibrated with the experimental $Q-V$ data from MFM based on 10 nm \HZO~for various $x$ (Fig. 3 (a)). Fig 4(a) shows the calibrated $h$ and $g_y$ parameters and the other parameters of the model are listed in Fig. 2(b).

The model successfully captures the evolution of the $Q-V$ response from a single FE loop to double AFE loops as $x$ increases from 0.5 to 1.0 (Fig. 3(a)).
At low $x$ ($x$ = 0.5--0.6), the $Q-V$ characteristic shows a single-loop response (for example, with one positive coercive voltage during the positive \Vapp~sweep).
Around $x$ = 0.7--0.8, the loop begins to pinch near \Vapp~$\approx 0$ V \cite{das_sub_2022,jung_review_2022}.
At this point, the coercive voltage starts to split from a single value into two values as shown in \cite{xu_pinched_2016}.
With further increase in $x$, the two coercive voltages separate further (one shifts toward positive bias and the other toward negative bias), producing distinct positive and negative coercive voltages and, consequently, a double-loop response \cite{randall_antiferroelectrics_2021}.
To quantify the loop ``separation'' from a single-loop to a double-loop response observed with increasing $x$, we introduce two characteristic voltages, $V_1$ and $V_2$ during the positive \Vapp~sweep.
$V_1$ is defined as the \Vapp~at which $P$ reversal initiates from single domain (SD) \textit{o}(--)-phase (only negative $P$ across \HZO).
Similarly, $V_2$ is the \Vapp~at which entire $P$ across \HZO~turns into positive $P$, i.e., SD \textit{o}(+)-phase.
The separation between $V_2$ and $V_1$, defined as $V_{2-1} \equiv V_2 - V_1$, therefore quantifies the $x$-dependent loop-separation in $Q-V$.
At low Zr concentration ($x$ = 0.5--0.6), $V_1$ and $V_2$ are closely spaced, yielding an effectively single-loop hysteresis.
With increasing $x$, $V_1$ shifts toward negative bias while $V_2$ shifts toward positive bias, increasing $V_{2-1}$ as shown in Fig. 4(b).
This progressive splitting results in the FE-to-AFE \textit{Q-V} transition as $x$ increases.

\begin{figure}[!t]
\centerline{\includegraphics[width=\columnwidth]{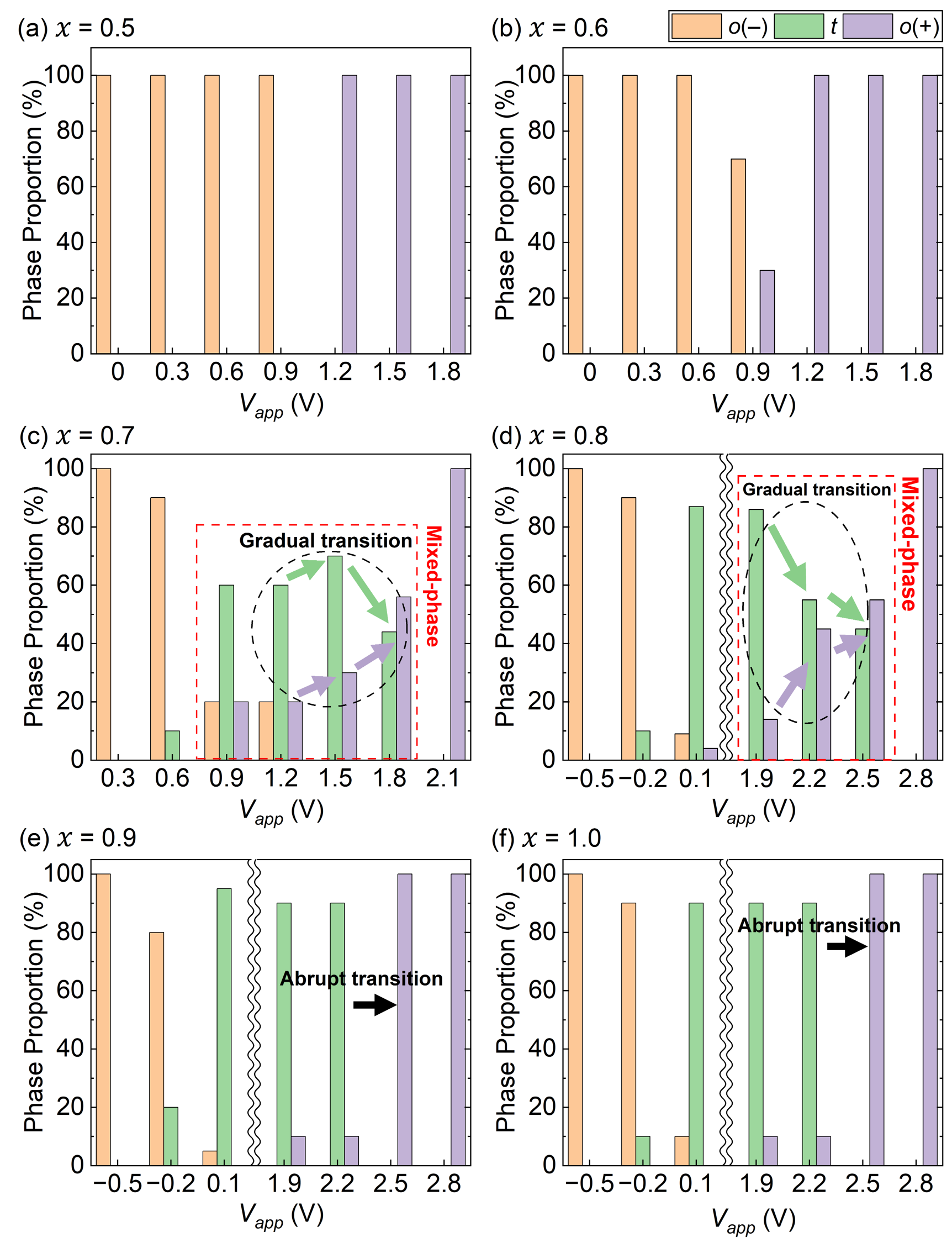}}
\caption{Phase (\textit{o}(--), \textit{t}, and \textit{o}(+)) composition for different applied voltages (\Vapp) during a positive sweep (from --4 V to +4 V) as a function of $x$. At $x$ = 0.7 and 0.8, \HZO~has heterogeneous (mixed) phase composition (red dashed box) and gradual phase transition (black dashed oval), which leads to a gradual $Q-V$ profile after the $V_1$ ($x$ = 0.7 and 0.8 in Fig. 3(a)). At low (= 0.5 and 0.6) and high $x$ (= 0.9 and 1.0), homogeneous phases and abrupt phase transitions are observed, resulting in abrupt $Q-V$.}
\label{fig6}
\end{figure}
\begin{figure*}[!t]
\includegraphics[width=\textwidth]{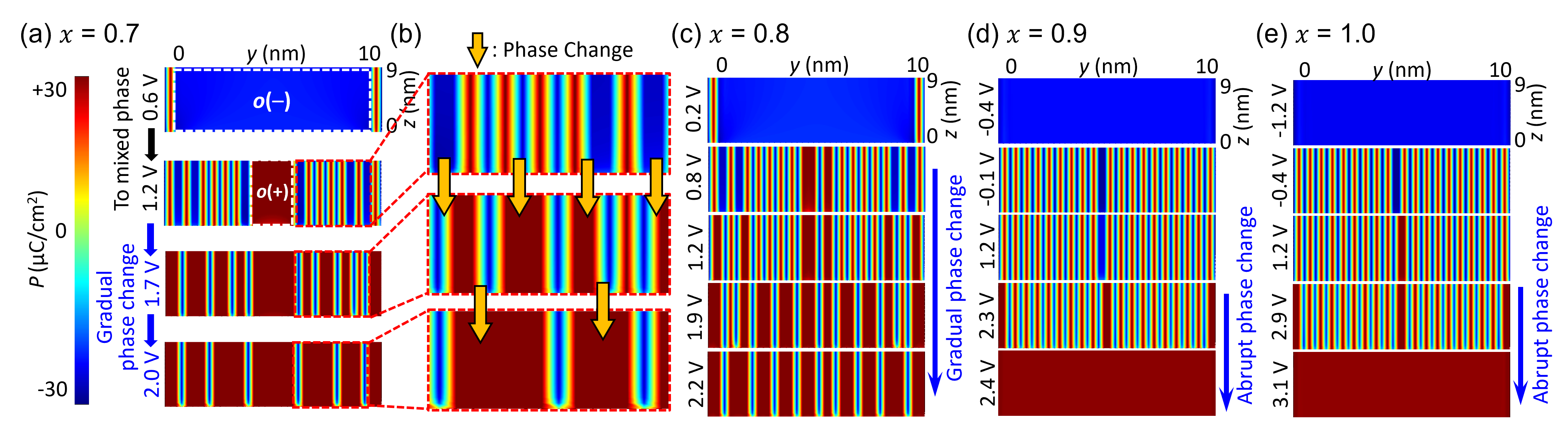}
\caption{Local $P$ maps of \HZO~for different \Vapp~at (a) $x$ = 0.7 (c) 0.8 (d) 0.9 and (e) 1.0. At $x$ = 0.7, the homogeneous \omp~domain (white dashed box) transitions to multiple phases between 0.6 and 1.2 V, followed by (b) gradual $P$ switching for \Vapp $>$ 1.2 V. Likewise, this mixed phase and gradual switching are observed at $x$ = 0.8. In contrast, $x$ = 0.9 and 1.0 display mostly homogeneous phase composition across the entire \Vapp~range}
\label{fig7}
\end{figure*}
\begin{figure}[!t]
\centerline{\includegraphics[width=\columnwidth]{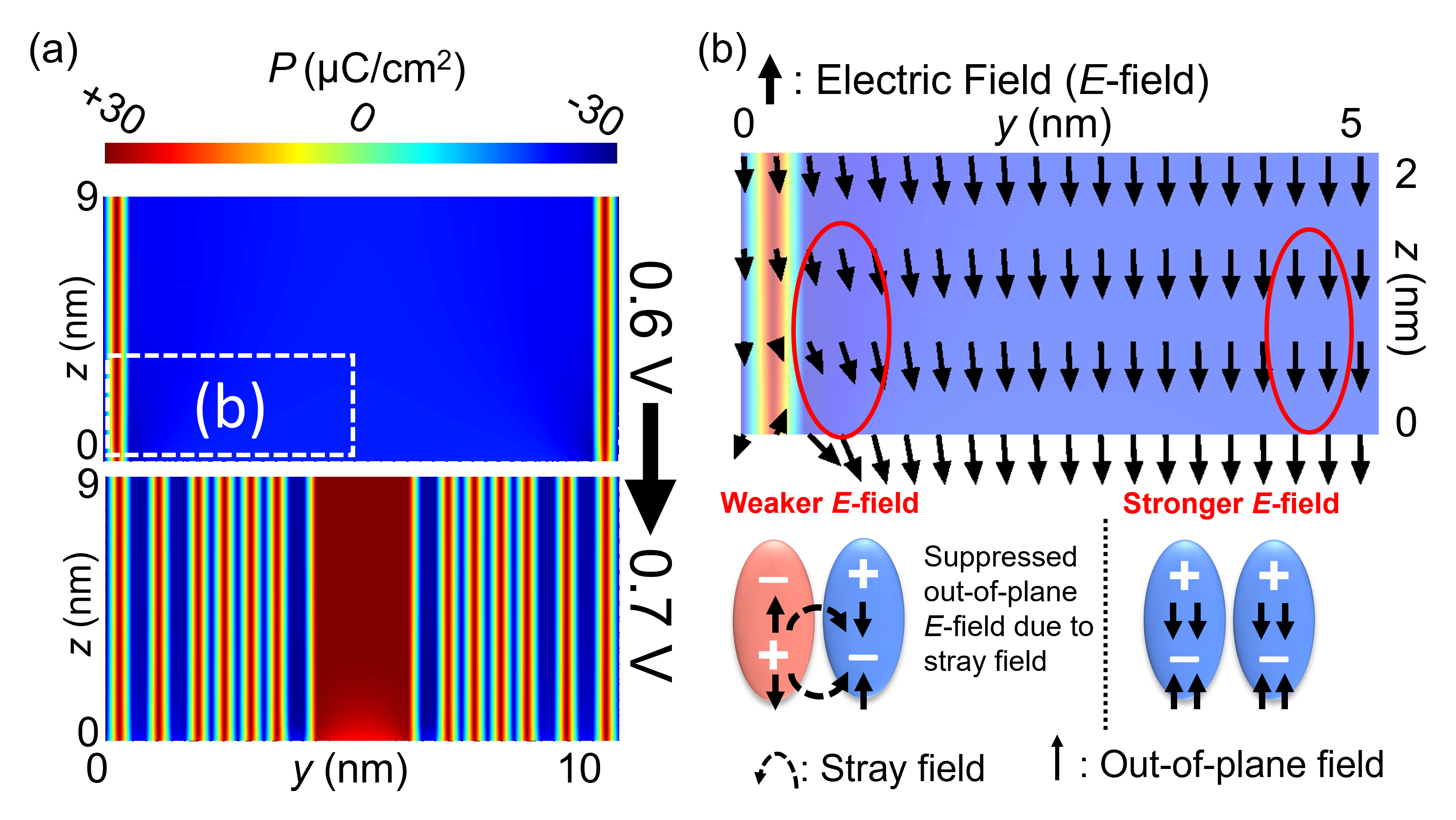}}
\caption{(a) Local $P$ maps of \HZO~layer with $x$ = 0.7 at \Vapp~= 0.6 to 0.7 V and (b) the \textit{E}-field distribution of the white dashed box in (a) (top). The spatial variation in the \textit{E}-field can be explained with the stray field effect due to opposite $P$ suppressing the out-of-plane \textit{E}-field (bottom). At 0.7 V, the homogeneous \textit{o}(--)-phase domain transitions to \textit{t} (near edge) and \opp~(bulk).}
\label{fig8}
\end{figure}
\section{Analysis of Zr Concentration Effects}
\subsection{\label{sec:level2}$x$-dependent Lattice Energy and its Effect on $Q-V$}
To interpret the $x$-dependent evolution of the $Q-V$ hysteresis, we first analyze the underlying \Utot~landscape of a single lattice.
This is because the phase preference and transition, which determine macroscopic $Q-V$ response, are governed primarily by (i) the relative energy levels of the competing phases (thermodynamic preference) and (ii) the energy barriers between them (kinetic transition dynamics) \cite{schroeder_fundamentals_2022}.
To this end, we compute 2-D \Utot~contour maps for different $x$ values using the calibrated $h$ and $g_y$ parameters in (2) and (3) as plotted in Fig. 3(b).
Starting from a substantially lower \Utot~for the \textit{o}-phase at low $x$ (= 0.5), the relative phase stability progressively reverses as the \Utot~of the \textit{t}-phase decreases while that of the \textit{o}-phase increases with increasing $x$.

Importantly, the applied \textit{E}-field transforms the energy profile through electrostatic coupling (see equation (4)) in a way that lowers the relative energy of the most preferred phase and accordingly changes the energy barriers.
A field-driven phase transition occurs under a particular \textit{E}-field condition where the corresponding barrier collapses \cite{reyes-lillo_antiferroelectricity_2014,qi_phase_2020}.
Because this critical \textit{E}-field is proportional to the energy barrier height (a larger \textit{E}-field is required to collapse a higher energy barrier), analyzing the $x$-dependence of the barrier is important to clarify the physical origin of the trends in $V_1$ and $V_2$. This is analogous to understanding the critical \textit{E}-field for phase transitions.

To that end, we obtain the energy barriers along the phase transition path connecting the \textit{o-} and \textit{t}-phase by projecting the path from the 2-D \Utot~contour map (Fig. 1(c)) onto a 1-D energy profile (Fig. 4(c)).
The \textit{o}-to-\textit{t} barrier is defined as the \Utot~difference between the \textit{o}-phase local minimum and the local maximum along the \textit{o} $\rightarrow$ \textit{t} path. The \textit{t}-to-\textit{o} barrier is defined in a similar manner (Fig. 4(c)).
Owing to the symmetry of \Utot~under $P$ reversal, the \textit{t} $\leftrightarrow$ \textit{o}(--) and \textit{t} $\leftrightarrow$ \textit{o}(+) transitions have identical energetics (i.e., the corresponding transition paths and barrier heights are the same).
Therefore, it is sufficient to analyze a single \textit{o} $\leftrightarrow$ \textit{t} path among the four phase transition paths in Fig. 1(c) to describe transitions among \textit{o}(--)-, \textit{t-}, and \opp s.
Both \textit{o}-to-\textit{t} and \textit{t}-to-\textit{o} barrier heights are summarized as a function of $x$ in Fig. 4(d).
According to Fig. 4(b) and (d), a similar trend is observed between \textit{o}-to-\textit{t} barrier and $V_1$  with respect to $x$. Further, \textit{t}-to-\textit{o} barrier and $V_{2-1}$ show similar trend.
In the following paragraph, we investigate this correlation by analyzing the field-driven transformation of the phase transition paths for low ($x$ = 0.5--0.6), intermediate ($x$ = 0.7--0.8), and high ($x$ = 0.9--1.0) Zr concentrations through \Utot~2-D contour maps (Fig. 5(a)--(c)) and 1-D projected phase transition paths (Fig. 5(d)).

At low concentration ($x$ = 0.5--0.6), FE behavior of $Q-V$ is attributed to the thermodynamic stability of the \textit{o}-phase, which is significantly more energetically favorable than the \textit{t}-phase at this composition (Fig. 3(b)).
This \textit{o}-phase preference persists even at non-zero \textit{E}-field conditions, where the stable $P$ sign is determined by \textit{E}-field sign (Fig. 5(a)).
Under a large negative \textit{E}-field ($E <<$ 0 in Fig. 5(a)), the \textit{o}(--)-phase is the most stable state with a huge energy barrier, resulting in a finite negative $Q$ at negative \Vapp~(Fig. 3(a)).
As $E$ is swept to the positive side, the \Utot~of the \textit{o}(--)-phase increases, progressively reducing the \textit{o}-to-t barrier.
Due to the large \textit{o}-to-\textit{t} barrier (Fig. 4(d)), sufficiently large \textit{E}-fields are needed to overcome it --- corresponding to large $V_1$ (Fig. 4(c)) --- thereby enabling phase transition out of the \textit{o}(--)-phase.
Meanwhile, \textit{t}-phase does not retain as a stable phase because of a marginal \textit{t}-to-\textit{o} barrier (Fig. 4(d)), which collapses even under a weak \textit{E}-field.
Therefore, under a positive \textit{E}-field, the \textit{o}(+)-phase remains the only stable phase that the \textit{o}(--)-phase can switch into ($E >0$ in Fig. 5(a)).
Consequently, the dominant transition is \textit{o}(--) $\rightarrow$ \textit{o}(+), manifested as a sharp change in $Q$ from a negative value to a positive value.
This brings $V_2$, where transition to \opp~completes, very close to $V_1$, yielding an almost zero $V_{2-1}$ (Fig. 4(b)). Thus, small $V_{2-1}$ can be thought of as a consequence of small \textit{t}-to-\textit{o} barrier. 
The same mechanism applies during the negative \Vapp~sweep due to the symmetry in \Utot, but with opposite polarity. 

At high Zr concentrations ($x$ = 0.9--1.0), the model accurately reproduces the AFE double-loop hysteresis, as shown in Fig. 3(a).
Starting from the stable \omp~state at large negative \Vapp~($E << 0$ in Fig. 5(b)), the \Utot~of \omp~rises and the non-polar \textit{t}-phase becomes increasingly favored as \Vapp~is swept toward positive bias ($E = 0$ and $> 0$ in Fig. 5(c)).
During this process, the \omp~can transition into other phases even before the \textit{E}-field reaches a large positive magnitude (unlike low $x$). This is because of the low \textit{o}-to-\textit{t} barrier, yielding less positive $V_1$ as shown in Fig. 4(b).
The \textit{t}-phase becomes the most stable phase at this point, which makes \textit{o}(--) $\rightarrow$ \textit{t} a major phase transition path.
Therefore, $Q-V$ curves display near-zero $Q$ after passing $V_1$ ($x$ = 0.9--1.0 in Fig. 3(a)).
After entering the \textit{t}-phase, the system remains there over an extended bias range because the \textit{t}-to-\textit{o} barrier is still large (Fig. 4(d)), even though \opp~is increasingly stabilized as the field increases (Fig. 5(c)).
Only when a substantially larger positive \textit{E}-field is applied, the \textit{t}-to-\textit{o} barrier collapses, enabling the \textit{t} $\rightarrow$ \opp~transition, thereby defining $V_2$.
Therefore, the large \textit{t}-phase barrier directly translates into a wide separation between $V_2$ and $V_1$ (i.e., a large $V_{2-1}$) (Fig. 4(d)).
Because the same transition occurs during the negative \Vapp~sweep with the opposite polarity, the $Q-V$ curves form a double loop with reversed orientation.

At intermediate concentrations ($x$ = 0.7--0.8), the model predicts a pinched $Q-V$ hysteresis loop (Fig. 3(a)) with $V_1$ and $V_{2-1}$ that fall between those of the low and high-$x$ (Fig. 4(b)).
This intermediate behavior is consistent with (i) a moderate \textit{o}-to-\textit{t} barrier (yielding an intermediate $V_1$) and (ii) \textit{t}-to-\textit{o} barrier (so the \textit{E}-field required to complete switching into the \opp~state, i.e., $V_{2-1}$, is also intermediate). This follows the same analysis of $E$-dependent barrier transformation that we presented for low and high $x$ (Fig. 5(b)).
Here, it is notable that the energy thresholds for the \textit{o}(--) $\rightarrow$ \textit{t} and \textit{t} $\rightarrow$ \textit{o}(+) transitions are almost the same in this $x$ range due to the comparable height of \textit{o}-to-\textit{t} and \textit{t}-to-\textit{o} barrier.
The nearly equal transition thresholds lead to strong competition between \textit{t}- and \opp~near the onset of switching from \omp, enabling mixed-phase configurations (details in later).
As a result, the $Q-V$ curves show gradual switching (even for a single grain considered in this analysis). 
This is a key characteristic signature of mixed-phase responses reported near this Zr concentration \cite{hyukpark_surface_2017,park_morphotropic_2018,ni_equivalent_2019}.
\begin{figure}[t]
\centerline{\includegraphics[width=\columnwidth]{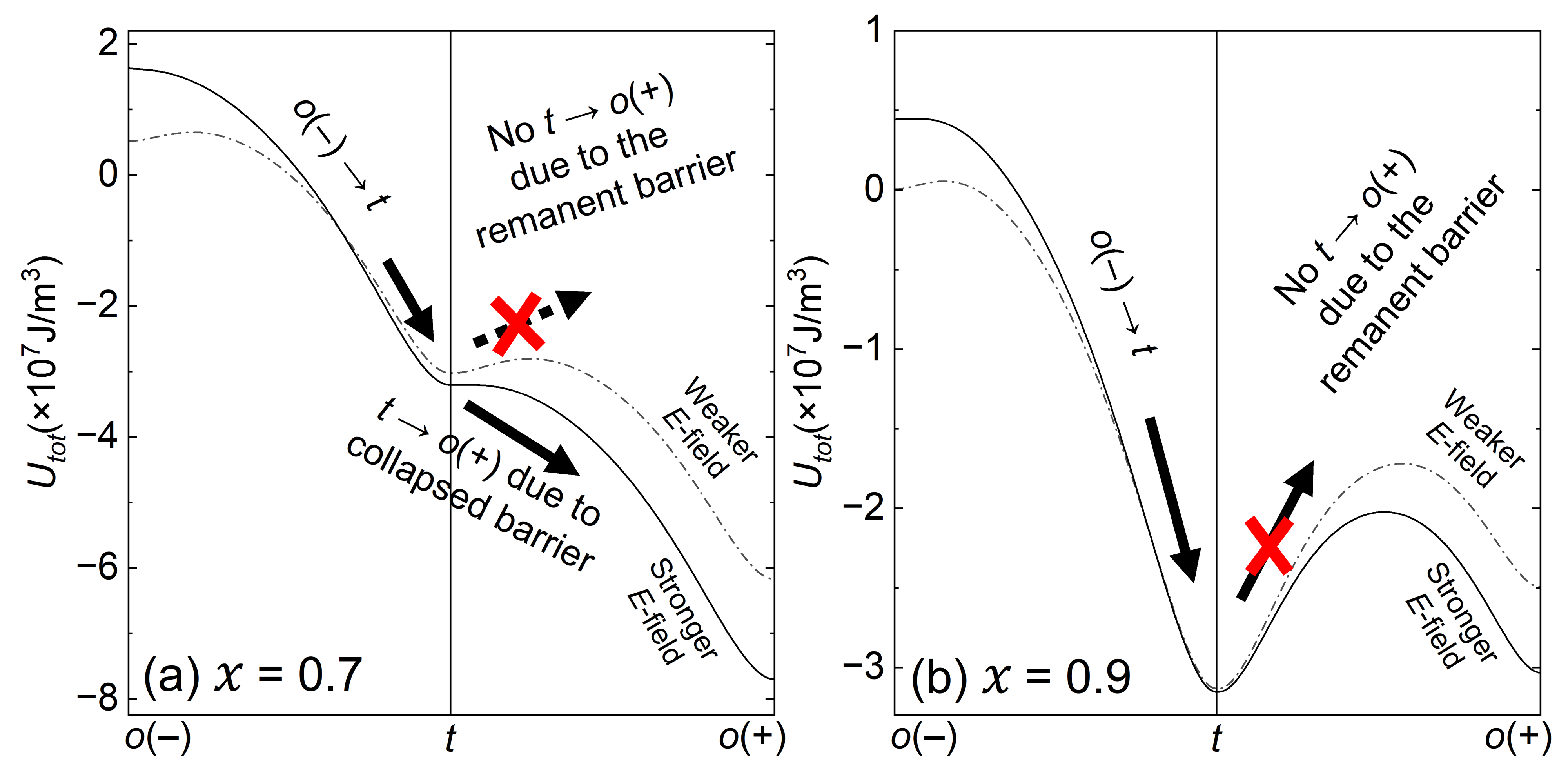}}
\caption{Phase transition plots from \textit{o}(--)- to \opp~through \textit{t}-phase when $x$ = (a) 0.7 and (b) 0.9 for different strengths of \textit{E}-field (solid lines: stronger and dashed lines: weaker \textit{E}-field)}
\label{fig9}
\end{figure}
\subsection{\label{sec:level2}Phase Composition During \Vapp~Sweep for Various $x$}

While the single-lattice \Utot~landscape analysis provides valuable insights into the thermodynamic preference and kinetic transition of \textit{o-} and \textit{t}-phases as a function of $x$, it is insufficient to fully capture the mixed-phase behavior (such as gradual $P$ switching), particularly prominent around $x$ = 0.7--0.8.
Therefore, it is necessary to investigate additional physical mechanisms beyond uniform lattice energetics.
To account for this, our phase-field framework extends the analysis to spatially resolved phase configurations by leveraging the simulated spatial $P$ map of the \HZO~domains (Fig. 2(c)).

Fig. 6 summarizes the simulated phase composition as a function of \Vapp~during positive \Vapp~sweep (from --4 V to +4 V) for various $x$ by plotting the phase proportion of \omp~(orange), \textit{t}-phase (green), and \opp~(purple).
These proportions are extracted from the simulated spatial $P$ maps by pairing adjacent sub-lattices into a single lattice and classifying each pair based on the relative sign of the two sub-lattice $P$ (Fig. 2(d)).
It provides a detailed analysis of the phase composition and \Vapp-driven phase transition for different $x$, linking the microscopic mechanism to the corresponding macroscopic $Q-V$ response.

Importantly, this analysis predicts the mixed \textit{o-} and \textit{t}-phase compositions and explains the gradual switching characteristics, which are prominent at the intermediate $x$ range ($x = 0.7$ and 0.8).
Fig. 6(c)--(d) exhibit an extended mixed-phase window (red dashed boxes) in which \textit{o}(--)-, \textit{t-}, and \opp~coexist after the \HZO~layer departs from the SD \omp~state.
Within this window, the phase fractions evolve gradually with increasing \Vapp~(black dashed ovals), consistent with gradual $Q-V$ responses beyond $V_1$ ($x$ = 0.7 and 0.8 in Fig. 3(b)).
This mixed-phase behavior is more pronounced at $x = 0.7$, which exhibits a wider voltage window of gradual transition. On the other hand, $x = 0.8$ shows a comparatively higher \textit{t}-phase fraction and a narrower window for gradual switching, which stems from decreased \textit{t}-phase \Utot~(increased stability) (see Fig. 3(b)).

In contrast, at low ($x$ = 0.5--0.6) and high ($x$ = 0.9--1.0) Zr concentrations,  \HZO~exhibits a predominantly homogeneous phase over most of the \Vapp~range.
At the low $x$, the homogeneous \omp~distribution rapidly turns into homogeneous \opp~within a narrow \Vapp~interval, consistent with a direct $P$ reversal within the \textit{o}-phase (\textit{o}(--) $\rightarrow$ \textit{o}(+)).
This abrupt conversion explains the sharp single-domain FE-like switching in a grain and the small separation between $V_1$ and $V_2$ in this regime.
For high Zr concentration ($x$ = 0.9--1.0), the \HZO~layer first transitions from the homogeneous \omp~state into a predominant \textit{t}-phase configuration, which persists over a wide \Vapp~range.
With further increase in \Vapp, the \textit{t}-phase fraction collapses abruptly, and the layer converts nearly entirely to \opp.
This sharp composition change corresponds to the field-induced \textit{t} $\rightarrow$ \textit{o}(+) transition that produces AFE double-loop response and a large separation between $V_1$ and $V_2$.
\begin{figure}[!t]
\centerline{\includegraphics[width=\columnwidth]{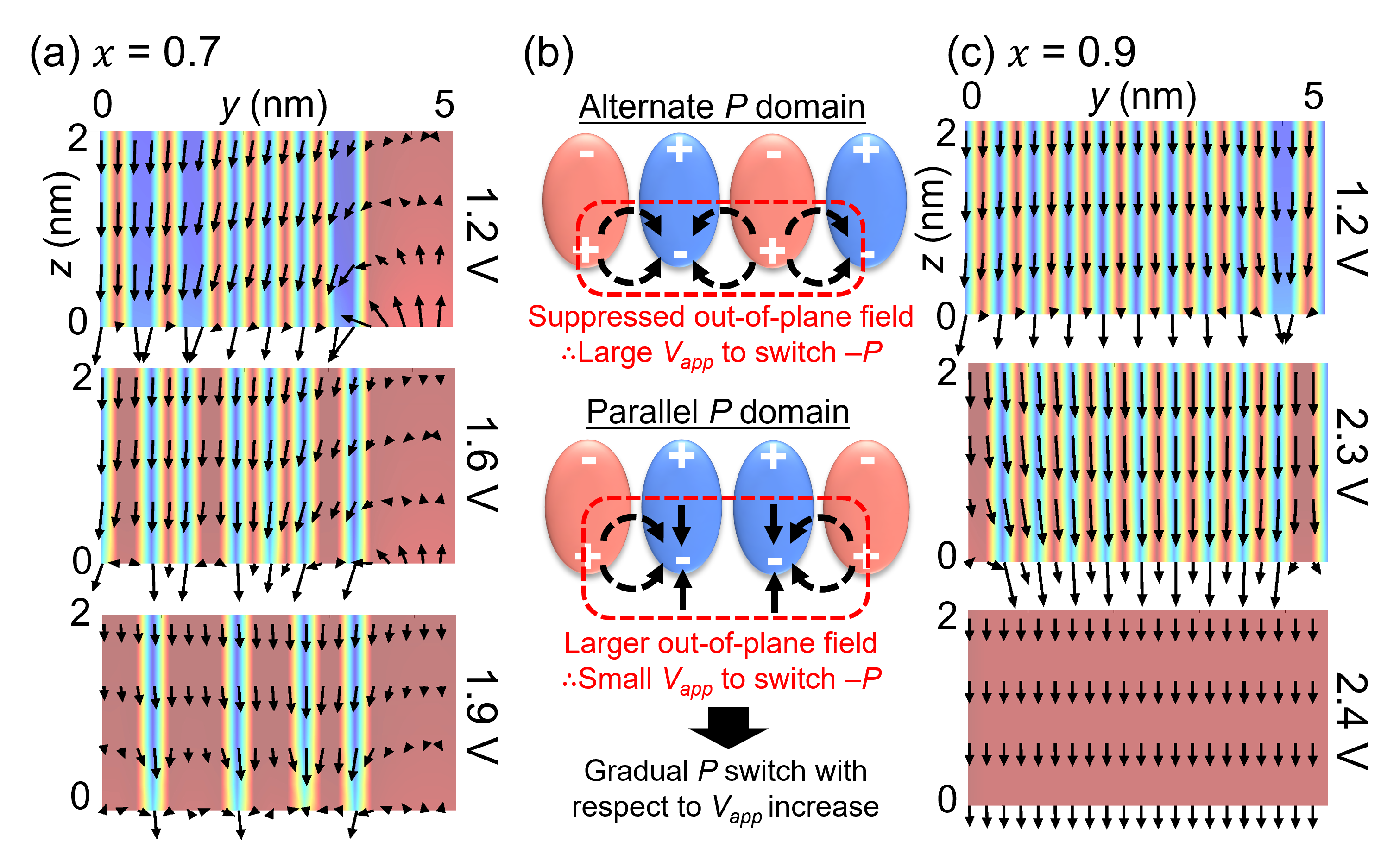}}
\caption{\textit{E}-field map across \HZO~layer for $x$ = (a) 0.7 and (c) 0.9. Due to the irregular $P$ pattern caused by mixed-phase, \textit{E}-fields become non-uniform. Thus, switching voltage becomes different for different locations as described in (b).}
\label{fig10}
\end{figure}
\subsection{\label{sec:level2}Spatial Analysis on $P$ and \textit{E}-Field Distributions}
To gain further insight into the phase distribution and switching dynamics---particularly the mixed-phase response and gradual transition at intermediate $x$---we analyze the spatial profiles of 
$P$ and \textit{E}-field across the \HZO~layer.
In particular, the simulated $P$ maps at $x = 0.7$ (Fig.~7(a)) directly visualize the emergence of mixed-phase domains during the \Vapp~sweep.
Starting from a nearly uniform \omp~state at 0.6 V (white dashed box in Fig. 7(a)), increasing \Vapp~drives the system into a heterogeneous configuration in which \textit{o}(--)-, \textit{t-}, and \opp~coexist (1.2 V in Fig. 7(a)).
Notably, a \opp~domain forms near the center of the layer, whereas the \textit{t}-phase forms near the edges.
With further increase in \Vapp, remaining \textit{t} and \omp~in the side region (red dashed box) progressively convert to \opp~(Fig. 7(b)).
This spatially staggered phase conversion provides a microscopic explanation for the gradual $Q-V$ evolution observed at $x$ = 0.7 (Fig. 3(b)) and for the comparable phase fractions extracted in Fig. 6(c).

For $x = 0.8$ (Fig.~7(c)), \HZO~exhibits a similar mixed-phase evolution after departing from \omp. 
However, the central \opp~region is reduced, and the surrounding \textit{t}-phase fraction remains larger.
This is consistent with the higher \textit{t}-phase proportion observed at $x = 0.8$ in Fig.~6(d) relative to $x = 0.7$.

The mixed-phase response and gradual switching originate from the interplay between (i) spatially non-uniform \textit{E}-field distributions induced by the multi-domain $P$ patterns and (ii) the $x$-dependent \Utot~landscape that sets the relative \textit{o}/\textit{t} stability and transition barriers.
Fig.~8(b) illustrates how $P$ patterns impact \textit{E}-field distribution, specifically the out-of-plane component that drives $P$ switching.
At the edges of the simulation domain (grain boundaries), each boundary sub-lattice interacts with only one neighboring sub-lattice (through gradient and interaction terms).
This local asymmetry drives the edge sub-lattice pair toward anti-parallel $P$, which indicates a stronger local \textit{t}-phase preference than in the bulk region. 
Note that in this case, the boundary sub-lattices become more negative and their neighboring sub-lattices correspondingly more positive because of $h > 0$.
When opposite $P$ are adjacent, as illustrated near the edges of the simulated region in Fig.~8(b), bound charges generate in-plane stray fields that suppress the out-of-plane \textit{E}-field component \cite{saha_ferroelectric_2020}.
In contrast, regions near the center are predominantly \textit{o}-phase (parallel $P$) due to sub-lattice interaction from both sides.
Therefore, these regions experience minimal stray-field screening, thus sustaining a stronger out-of-plane \textit{E}-field.

This spatial non-uniformity in \textit{E}-field directly impacts the phase transition pathway.
At intermediate $x$, the energy barriers associated with \textit{o}(--) $\rightarrow$ \textit{t} and \textit{t} $\rightarrow$ \opp~transitions can collapse almost simultaneously when the local out-of-plane \textit{E}-field reaches the switching threshold (the solid line in Fig. 9(a)), due to the comparable magnitude of the \textit{o}-to-\textit{t} and \textit{t}-to-\textit{o} barriers (Fig. 4(d)).
As a result, \omp~domain in the center turns into \opp~domain because of stronger \textit{E}-field promoting \textit{o}(--) $\rightarrow$ \textit{o}(+) transition (Fig. 8(b)).
On the other hand, weaker-\textit{E}-field at the edges (Fig. 8(b)) mainly remains \textit{t}-phase because of the remaining barrier between \textit{t} and \opp~blocking the phase transition to \opp~(the dashed dot line in Fig. 9(a)).
In this regime, the comparable \textit{o}-to-\textit{t} and \textit{t}-to-\textit{o} barrier heights make the phase transition path highly sensitive to local \textit{E}-field variations.
Consequently, spatial \textit{E}-field variation directly leads to heterogeneous phase composition by impacting the phase transition paths.

In contrast, \HZO~with high $x$ (= 0.9--1.0) features homogeneous \textit{t}-phase distribution and abrupt phase change (Fig. 7(d)--(e)).
This is because the influence of the non-uniform \textit{E}-field distribution becomes marginal.
Here, the \textit{t} $\rightarrow$ \textit{o}(+) barrier (\textit{t}-to-\textit{o}) is substantially higher than the \textit{o}(--) $\rightarrow$ \textit{t} barrier (\textit{o}-to-\textit{t}) (Fig.~4(d)).
Thus, even after the \textit{o}(--) $\rightarrow$ \textit{t} barrier collapses, a large residual barrier continues to block the \textit{t} $\rightarrow$ \textit{o}(+) conversion.
As a result, even the stronger \textit{E}-field (dashed line in Fig. 9(b)) is insufficient to overcome this high energy barrier and induce a phase transition.
Therefore, the system remains in a homogeneous \textit{t}-phase until the applied bias is strong enough to simultaneously overcome the energy barrier across the entire film.

Now, let us turn our attention to gradual switching at the intermediate $x$.
The gradual switching is mainly governed by the coupled effects of (i) domain wall (DW)-induced stray fields, which create a spatially non-uniform out-of-plane \textit{E}-field and (ii) the near-equal \Utot~barriers between competing and \textit{t-} and \opp s.
The coexistence of \textit{o}(--)-, \textit{t-}, and \opp s generates an irregular $P$ domain pattern (Fig. 10(a)), which consists of both alternating and parallel $P$ domains (Fig. 10(b)).
This irregularity leads to significant \textit{E}-field spatial variations (Fig. 10(a)), meaning \HZO~sub-lattices experience different switching forces depending on their location.
In addition, the pronounced sensitivity of the \Utot~landscape to local \textit{E}-field variations magnifies these spatial differences, causing certain regions to cross the transition barrier earlier while others remain delayed.
Regions subjected to stronger fields switch earlier by overcoming a lowered local \textit{t} $\rightarrow$ \textit{o}(+) energy barrier, whereas those under weaker fields require a higher \Vapp~to overcome their local energy barriers (Fig. 9(a)).
Consequently, $P$ switching occurs in a spatially staggered manner across the grain, leading to a gradual macroscopic \textit{Q-V} response rather than an abrupt transition.

In contrast, at high $x$ (= 0.9--1.0), the phase homogeneity in \HZO~leads to more uniform \textit{E}-fields due to regular $P$ domain patterns (Fig. 10(c)). 
On top of that, the negligible influence of local \textit{E}-field variation due to the high energy barrier for the \textit{t} $\rightarrow$ \textit{o}(+) transition (Fig. 9(b)) suppresses staggered $P$ switching. When the applied bias is strong enough to overcome the energy barrier, the entire film switches (Fig. 10(c)).
This abrupt and collective transition from a homogeneous \textit{t}-phase to an SD \textit{o}(+)-phase leads to sharper switching, as observed in \textit{Q-V} hysteresis for $x$ = 0.9--1.0 in Fig. 3(b).

\section{Conclusion}
We developed a self-consistent sub-lattice phase-field model to investigate Zr concentration ($x$)-dependent phase composition and polarization ($P$) switching in \HZO~capacitors.
The model reproduces the experimentally observed evolution of $Q-V$ hysteresis with $x$ and provides an interpretation of the FE-to-AFE crossover from both energetics and spatial perspectives.

The sub-lattice formulation captures the $x$-dependent \Utot~landscape of a single lattice, clarifying how thermodynamic stability and kinetic transition barriers determine switching voltages.
By incorporating the model within a phase-field framework, we provide insights into the spatial distributions of $P$ and \textit{E}-field, enabling mixed-phase and MD configurations to emerge naturally.

At low Zr concentration ($x$ = 0.5--0.6), the \textit{o}-phase is thermodynamically favored, resulting in predominantly homogeneous \textit{o}(--) $\rightarrow$ \textit{o}(+) reversal in the forward path and sharp FE switching.
At high Zr concentration ($x\geq 0.9$), the \textit{t}-phase is stabilized, producing a largely homogeneous AFE-like response characterized by an abrupt, field-induced \textit{t} $\rightarrow$ \textit{o}(+) transition in the forward path.
At intermediate concentrations ($x$ = 0.7--0.8), comparable \textit{o} $\leftrightarrow$ \textit{t} barrier heights and multi-domain $P$-induced stray fields, create a significantly non-uniform local \textit{E}-field.
This yields heterogeneous mixed-phase configurations and spatially staggered $P$ switching, which manifests as a gradual switching in the macroscopic $Q-V$ characteristics.

\begin{acknowledgments}
The authors acknowledge Revanth Koduru (Purdue) for his help with phase-field modeling and Prof. Suman Datta (Georgia Tech) for experiments in \cite{saha_microscopic_2019}.
\end{acknowledgments}

\section*{Data Availability Statement}
The data that support the findings of this study are available from the corresponding author upon reasonable request.

\bibliography{aipsamp}

\end{document}